\documentclass[a4paper,11pt]{article}
\pdfoutput=1

\usepackage{jheppub} 
\usepackage{graphicx,color}
\usepackage{enumitem}
\usepackage{braket}
\usepackage{slashed}
\usepackage[utf8]{inputenc}
\usepackage{hyperref}
\usepackage{float}
\usepackage{caption}
\usepackage{subfigure}
\usepackage{makecell}
\usepackage{epsfig}
\usepackage{amsmath,amssymb}
\usepackage{hyperref}
\hypersetup{
    colorlinks=flase,
    linkcolor=black,
    filecolor=magenta,      
    urlcolor=blue,
}

\hypersetup{
	colorlinks,
	citecolor=black,
	filecolor=black,
	linkcolor=blue,
	urlcolor=black
}

\newcommand{\bea}{\begin{eqnarray}}
\newcommand{\eea}{\end{eqnarray}}
\newcommand{\be}{\begin{equation}}
\newcommand{\ee}{\end{equation}}
\newcommand{\ba}{\begin{align}}
\newcommand{\ea}{\end{align}}
\newcommand{\ben}{\begin{enumerate}}
\newcommand{\een}{\end{enumerate}}
\newcommand{\bi}{\begin{itemize}}
\newcommand{\ei}{\end{itemize}}
\newcommand{\comments}[1]{}

\def\bel#1{\begin{equation} \label{#1}}

\title{Out of the Swampland with Multifield Quintessence?}

\author[a]{Michele Cicoli,}
\author[b]{Giuseppe Dibitetto}
\author[a]{and Francisco G. Pedro}
\affiliation[a]{Dipartimento di Fisica e Astronomia, Universit\`a di Bologna, via Irnerio 46, 40126 Bologna, Italy and INFN, Sezione di Bologna, viale Berti Pichat 6/2, 40127 Bologna, Italy}
\affiliation[b]{Dipartimento di Fisica e Astronomia, Universit\`a di Padova, via Marzolo 8, 35131 Padova, Italy and INFN, Sezione di Padova, via Marzolo 8, 35131 Padova, Italy}

\emailAdd{michele.cicoli@unibo.it}
\emailAdd{giuseppe.dibitetto@unipd.it}
\emailAdd{francisco.soares@unibo.it}

\abstract{Multifield models with a curved field space have already been shown to be able to provide viable quintessence models for steep potentials that satisfy swampland bounds. The simplest dynamical systems of this type are obtained by coupling Einstein gravity to two scalar fields with a curved field space. In this paper we study the stability properties of the non-trivial fixed points of this dynamical system for a general functional dependence of the kinetic coupling function and the scalar potential. We find the existence of non-geodesic trajectories with a sharp turning rate in field space which can give rise to late-time cosmic acceleration with no need for flat potentials. In particular, we discuss the properties of the phase diagram of the system and the corresponding time evolution when varying the functional dependence of the kinetic coupling. Interestingly, upon properly tuning the initial conditions of the field values, we find trajectories that can describe the current state of the universe. This could represent a promising avenue to build viable quintessence models out of the swampland if they could be consistently embedded in explicit string constructions.}

\begin{document} 
\maketitle
\flushbottom

\section{Introduction}
\label{Intro}

Shortly before the turn of the millennium we learned that the universe is currently undergoing a phase of accelerated expansion \cite{Riess:1998cb,Jaffe:2000tx,Tegmark:2003ud}, which reveals the existence of what is usually referred to as dark energy. The simplest cosmological model successfully accounting for this late-time dynamics is the so-called $\Lambda$CDM concordance model, in which a positive cosmological constant is responsible for dark energy. Despite its compelling experimental evidence though, this conclusion raises a number of theoretical issues which are still open after more than two decades.

The aforementioned issues are all mainly connected with the fact that a spacetime with positive cosmological constant turns out to possess a cosmological horizon when described within its static patch. Hence no obvious prescription seems to exist in order to consistently define a unitary quantum field theory on such a background. This issue has so far represented an obstruction in understanding an accelerated universe within a UV complete quantum theory by adopting a bottom-up approach.

On the other hand, from a top-down perspective, high-energy theoretical physicists such as string theorists have been proposing models that could explain how a low-energy effective de Sitter (dS) universe could possibly emerge from a suitable compactification of string theory. Though several types of well-motivated constructions are currently available on the market \cite{Kachru:2003aw,Westphal:2006tn,Cicoli:2012fh,Cicoli:2013cha,Cicoli:2015ylx,Gallego:2017dvd,Heckman:2018mxl},  it is often argued that they all raise a number of open issues concerning the use of exotic ingredients or the absence of control on perturbative and non-perturbative corrections given that supersymmetry is necessarily broken if $V>0$. 

While the issue is far from settled in general (see \cite{Cicoli:2018kdo} for a recent critical discussion of these issues in string compactifications), the many subtleties and obstacles encountered along the way motivate one to try and adopt a totally different approach to the problem. Since the original work of \cite{Obied:2018sgi,Garg:2018reu}, people started conjecturing the idea that the absence of dS vacua should actually be a desirable feature for any sensible candidate theory of quantum gravity \cite{Brennan:2017rbf,Danielsson:2018ztv}. This conjecture was then generalized and extended in \cite{Ooguri:2018wrx,Andriot:2018wzk}. However, in its most recent form, the dS swampland conjecture can be formulated in terms of a bound on the flatness of a scalar potential in the regions of field space where it becomes positive. If this conjecture were correct, not only would dS vacua be absent in our theory of quantum gravity, but also effectively single-field slow-roll quintessence models \cite{Kaloper:2005aj,Kaloper:2008qs,Panda:2010uq,Cicoli:2012tz,Blaback:2013fca,DAmico:2018mnx} would be ruled out of the landscape. The only models which survive are those based on steep potentials. However all known string quintessence constructions which are compatible with the swampland bound are in disagreement with observations \cite{Akrami:2018ylq}. 

From the phenomenological point of view, only a small class of models seems to be consistent with both swampland and observational constraints \cite{Scolnic:2017caz} and it is characterized by exponential potentials of the form $V=V_0 \,e^{-c \phi/M_\mathrm{Pl}}$ with $c\lesssim 0.6$ but still for $\mathcal{O}(1)$ values of the parameter $c$ \cite{Agrawal:2018own}. So far, no explicit string construction with this property has been derived. Moreover such a quintessence model would inevitably share the same control issues of existing dS vacua \cite{Cicoli:2018kdo, Akrami:2018ylq} together with additional problems typical of quintessence scenarios like those associated with the non-observation of fifth forces \cite{Acharya:2018deu}. 

Given that, from the theoretical point of view, quintessence models do not seem more robust than dS vacua, and the existing tension between the swampland bound and cosmological observations, it might well be that the dS swampland conjecture is too restrictive. In this work, we further investigate this possibility by looking for models of dark energy sourced by scalar fields in ways consistent with the dS swampland conjecture. Our starting point will be the analysis carried out in \cite{Cicoli:2020cfj}, which is based on the simple but crucial observation \cite{Brown:2017osf, Achucarro:2018vey,Christodoulidis:2019jsx} that field space curvature may in principle allow the existence of trajectories displaying arbitrarily many efoldings of cosmic acceleration even in presence of steep scalar potentials. The key to these genuinely multifield realizations of quintessence is non-geodesic motion in field space.\footnote{Our use of the term `non-geodesic' follows the conventions of the multifield inflation literature where `geodesic' trajectories are those following the gradient of the potential, while `non-geodesic' trajectories are characterized by a non-vanishing turning rate.} In this paper we extend our analysis to include more general functional dependence of the field space curvature. This will give rise to novel explicit examples where a suitably tuned initial condition for the fields yields accelerated trajectories with the relevant cosmological parameters compatible with those observed today.\footnote{See also \cite{Brahma:2019kch, Brahma:2020eqd} for an alternative approach where specific models can satisfy both observational and swampland constraints via the inclusion of higher derivative interactions in the effective action.}

In this paper we shall take a phenomenological point of view and just provide examples of multifield non-geodesic dynamics which can realize quintessence for steep potentials in agreement with the dS swampland conjecture. This represents however only the first step to test the robustness of this conjecture. The next crucial step would instead involve the explicit realization of consistent string models which can reproduce the promising phenomenological features described in this paper. A positive answer to this question would provide a strong support in favor of the dS swampland conjecture, especially if this kind of models would emerge naturally in string compactifications. A difficulty in embedding these multifield quintessence models in string theory would instead be a severe evidence against the dS swampland conjecture. 

Notice that the existence of these new accelerating solutions in late time cosmology would still be interesting \textit{per se}, even if the dS swampland conjecture turned out to be wrong, since they would provide a way to obtain viable quintessence models for steep potentials based on non-geodesic motion in a curved multifield space. This property is particularly important when quantum corrections are properly taken into account since in general they tend to spoil the flatness of the underlying potential.

The paper is organized as follows. In Sec. \ref{Sec2} we introduce the general setup, discuss the different physical properties of the various fixed points of the system and the phase diagram for generic kinetic coupling. In Sec. \ref{Sec3} we analyze how a quintessence model based on genuinely multifield dynamics and non-geodesic field space motion can be consistent with the bounds on the flatness of the scalar potential advocated by the dS swampland conjecture. In Sec. \ref{Sec4} and \ref{Sec5} we go through explicit examples of models with non-trivial two-field dynamics, with growing and decaying kinetic coupling, respectively. In Sec. \ref{Sec6} we briefly discuss the phenomenological relevance of power-law kinetic couplings and scalar potentials, thus abandoning exponential functional dependence in both. A brief summary of results and a discussion are presented in Sec. \ref{Sec7}.

\section{Overview of two-field dynamics with curved field space}
\label{Sec2}

Let us consider the effective theory of Einstein gravity coupled to two scalar fields named $\left(\phi_1,\phi_2\right)$, described by the following action
\be
S[g_{\mu,\nu},\phi] \, = \, \int d^4x \sqrt{-g} \, \left(\frac{M_{\mathrm{Pl}}^2}{2} \, \mathcal{R} \, - \, \frac{1}{2} \left(\partial\phi_1\right)^2 \, - \, \frac{1}{2} f(\phi_1)^2\left(\partial\phi_2\right)^2\, - \, V(\phi_1)\right) \ ,
\label{eq:S}
\ee
where the arbitrary function $f(\phi_1)$ specifies the scalar kinetic coupling. Our aim is to study the dynamics of the above system on a background geometry of the FRW type, i.e.
\be
ds_4^2 \, = \, - dt^2 \, + \, a(t)^2\,ds_{\mathbb{R}^3}^2  \qquad \textrm{ and } \qquad \phi_i \, = \, \phi_i(t) \ .
\ee
The equations of motion specified on the above \emph{ansatz} read
\be
\left\{
\begin{array}{ll}
\ddot{\phi}_1+3H\dot{\phi}_1-f\,f_1 \dot{\phi}_2^2+V_1=0 \ ,\\[5pt]
\ddot{\phi}_2+3H\dot{\phi}_2+2\frac{f_1}{f}\dot{\phi}_2\dot{\phi}_1=0 \ ,
\end{array}\right.
\ee
where $\cdot \equiv \frac{d}{dt}$, $H \equiv \frac{\dot a}{a}$ and we furthermore use the notation $f_1=\partial f/\partial \phi_1$ and similarly for $V$.
In the late universe, in the presence of a barotropic fluid with pressure $p_b=\omega_b \rho_b$ that evolves according to the continuity equation
\be
\dot{\rho}_b=-3 H (1+\omega_b) \rho_b\ ,
\ee
the Einstein equations imply the following form of the Friedman equation
\be
H^2=\frac{1}{3 M_{\mathrm{Pl}}^2}\left(\frac{\dot{\phi}_1^2}{2}+\frac{f^2}{2}\dot{\phi}_2^2+V+\rho_b\right) \ .
\label{eq:Fried}
\ee
Let us also define the kinetic coupling
\be
k_1(\phi_1)\equiv-M_{\mathrm{Pl}}\,\frac{f_1}{f} \ ,
\label{eq:k1Def}
\ee
and
\be
k_2(\phi_1)\equiv-M_{\mathrm{Pl}}\,\frac{V_1}{V}\ .
\label{eq:k2Def}
\ee

Note that, in terms of the above notation, the case of constant $k_1$ and $k_2$ has been analysed in \cite{Cicoli:2020cfj} and corresponds to fixing $V$ and $f$ to have exponential form. There we have shown that the inclusion of a kinetically coupled massless scalar enriched the quintessential dynamics by generating a novel fixed point that can help one account for the observed acceleration of the universe in accordance with earlier works \cite{Sonner:2006yn, Russo:2018akp}. Interestingly this new regime allows for accelerating solutions even when the scalar potential is steep ($k_2\gg1$). 

Following \cite{Copeland:1997et} it turns out to be useful to define the dimensionless variables
\be
x_1\equiv \frac{\dot{\phi}_1}{\sqrt{6} H M_{\mathrm{Pl}}}\ ,\quad x_2\equiv \frac{f \dot{\phi}_2}{\sqrt{6} H M_{\mathrm{Pl}}}\ ,\quad y_1\equiv \frac{\sqrt{V}}{\sqrt{3} H M_{\mathrm{Pl}}} \ ,
\ee
which allow us to rewrite the dynamics of the system as an autonomous system
 \be
 x_1'= 3 x_1 (x_1^2+ x_2^2-1 ) + \sqrt{\frac{3}{2}} (-2 k_1 x_2^2 + k_2 y_1^2) - \frac{3}{2} \gamma x_1 ( x_1^2 + x_2^2 + y_1^2-1 )\ ,
 \label{eq:dx1}
\ee
\be
 x_2'=3 x_2\left(x_1^2+x_2^2-1\right) +\sqrt{6} k_1 x_1 x_2 -\frac{3}{2} \gamma  x_2 \left(x_1^2+x_2^2+y_1^2-1\right)\ ,
  \label{eq:dx2}
 \ee
 \be
 y_1'= -\sqrt{\frac{3}{2}} k_2 x_1 y_1-\frac{3}{2} \gamma  y_1 \left(x_1^2+x_2^2+y_1^2-1\right)+3 y_1 \left(x_1^2+x_2^2\right)\ ,
  \label{eq:dy1}
\ee
 where $' \equiv \frac{d}{d \ln a}$.  Given that in the general case $k_1$ and $k_2$ are field dependent, these are to be supplemented by
\be
\phi_1'= 6 x_1 M_{\mathrm{Pl}}\ ,
\ee
or alternatively by one of the following equations for the time evolution of the parameters $k_1$ and $k_2$
 \be
 k_i'= \sqrt{6} k_i^2 x_1 \left(1-\Gamma_i\right) \ ,
 \label{eq:kprime}
 \ee
where $i=1,2$,  $\Gamma_1\equiv \frac{f_{11} f}{f_1^2}$ and $\Gamma_2\equiv \frac{V_{11} V}{V_1^2}$. We note that the field space curvature can be written as
\be
\mathbb{R}=\frac{f_1^2-2 f f_{11}}{2 f^2}=\frac{k_1^2}{2}-\Gamma_1\ .
\ee
The dynamics of the system can equivalently be recast in terms of the late time observables 
\be
\omega_\phi=\frac{p_\phi}{\rho_\phi}=\frac{x_1^2+x_2^2-y_1^2}{x_1^2+x_2^2+y_1^2}\ ,
\label{eq:omega}
\ee
and
\be
\Omega_\phi=x_1^2+x_2^2+y_1^2 \ .
\label{eq:Omega}
\ee
The evolution of these quantities follows from
\be
\Omega_\phi'=-3 (\Omega_\phi -1) \Omega_\phi  (\omega_b -\omega_\phi)\ ,
\ee
and
\be
\omega_\phi'=(\omega_\phi -1) \left(-k_2 \sqrt{3(\omega_\phi +1)\Omega_\phi -6 x_2^2}+3(1+ \omega_\phi) \right)\ .
\ee

Instantaneous fixed points are solutions to $x_1'=x_2'=y_1'=0$. For $\Gamma_1, \Gamma_2\neq 1$ these are not actual fixed points of the system, but are nonetheless useful for describing the dynamics of the system in a regime where the corresponding energy densities evolve faster than the couplings $k_1$ and $k_2$. There turn out to exist \emph{six} instantaneous fixed points, whose different physical properties were already studied in \cite{Cicoli:2020cfj}. They are again described below, for the sake of clarity.

\begin{itemize}
\item \textbf{Kinetic dominated fixed points:} $\mathcal{K_\pm}$\\
These fixed points correspond to kinetic domination. They are located at
\be
(x_{1},x_{2},y_{1})|_c=(\pm 1, 0,0)\ ,
\ee
and therefore, through Eqs. \eqref{eq:omega} and \eqref{eq:Omega}, $\Omega_\phi=1$ and $\omega_\phi=1$. These are always unstable nodes.
 
\item \textbf{Fluid dominated fixed point:} $\mathcal{F}$\\
This fixed point is located at 
\be
 (x_{1},x_{2},y_{1})|_c=(0, 0,0)\ .
 \ee
The scalar sector does not contribute to the energy density: $\Omega_\phi=0$. This fixed point turns out to be a saddle for $-1<\omega_b<1$.\\
 
\item \textbf{Scaling solution:} $\mathcal{S}$\\
At such a point the scalar field mimics the barotropic fluid and constitutes a subdominant component of the overall energy density. It is located at 
\be
(x_{1},x_{2},y_{1})|_c=\left(\frac{\sqrt{3/2}(\omega_b-1)}{k_2}, 0,\frac{\sqrt{3/2}\sqrt{1-\omega_b^2}}{k_2}\right) \ ,
\ee
with $\omega_\phi=\omega_b$ and $\Omega_\phi=\frac{3(1+\omega_b)}{k_2^2}$. The scaling fixed point is stable for $k_2>\sqrt{3}$ and $k_2>2 k_1$.\\

\item \textbf{Geodesic fixed point:} $\mathcal{G}$\\
At $\mathcal{G}$, $\phi_1$ slow-rolls down the slope of its potential, while $\phi_2$ remains frozen. It is located at
\be
(x_{1},x_{2},y_{1})|_c=\left(\frac{k_2}{\sqrt{6}},0, \sqrt{1-\frac{k_2^2}{6}}\right) \ ,
\ee
and is stable in the regime $k_2<\sqrt{3}$ and $0<k_2<\sqrt{k_1^2+6}-k_1$ and $k_1>0$ or $k_2>-\sqrt{3}$ and $-\sqrt{k_1^2+6}+k_1<k_2<0$ and $k_1<0$. In this fixed point $\omega_\phi=-1+\frac{k_2^2}{3}$ and $\Omega_\phi=1$.
 
\item \textbf{Non-geodesic fixed points:} $\mathcal{NG}$\\
At these fixed points  $\phi_2$ is dragged by the velocity of $\phi_1$. They are located at
\be
 (x_{1},x_{2},y_{1})|_c=\left( \frac{\sqrt{6}}{2k_1+k_2},\pm\frac{\sqrt{k_2^2+2 k_1 k_2 -6}}{|2 k_1+k_2|},\sqrt{\frac{2 k_1}{2 k_1+k_2}}\right) \ ,
\ee
and are stable in the region $\sqrt{k_1^2+6}-k_1<k_2<2 k_1$ and $k_2>0$ and $2 k_1<k_2<\sqrt{k_1^2+6}-k_1$ and $k_2<0$. In this fixed point $\omega_\phi=\frac{k_2-2k_1}{k_2+2k_1}$ and $\Omega_\phi=1$. 
\end{itemize}

\begin{figure}[!h]
  \centering
    \includegraphics[width= 0.60 \textwidth]{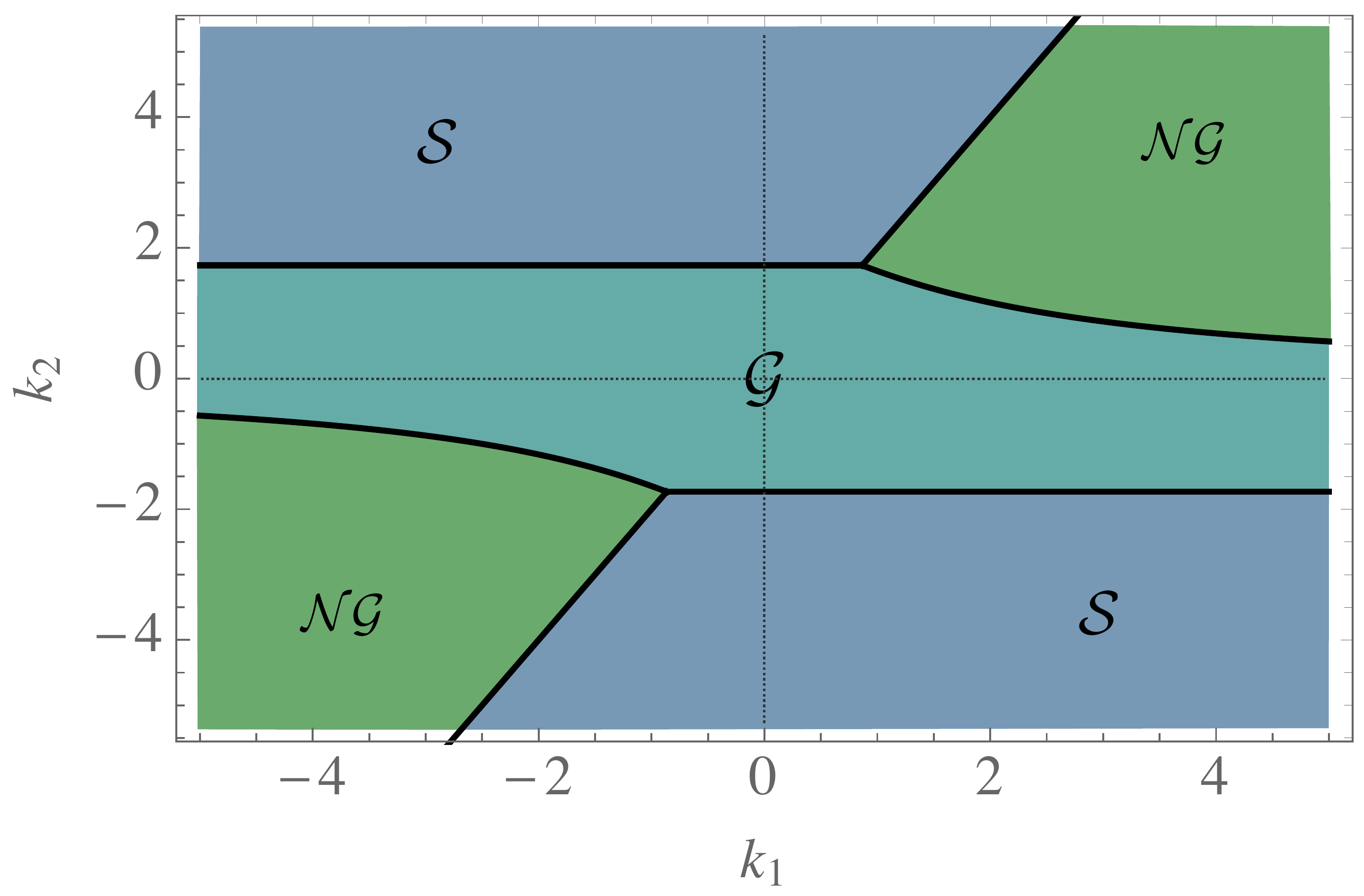}
  \caption{Full stability diagram for $\omega_b=0$.}
\label{fig:phaseDiag}
\end{figure}

In Fig. \ref{fig:phaseDiag} we plot the stability diagram for the $\mathcal{S},\mathcal{G}$ and $\mathcal{NG}$ fixed points.

In this paper we extend the analysis performed in \cite{Cicoli:2020cfj}, where both $k_1$ and $k_2$ were constant by considering non-trivial functional dependence in the $k_i$'s. For most of the paper we shall however restrict ourselves to situations where $k_2$ is still kept constant, while only relaxing the assumption of constant $k_1$ in our search for phenomenologically viable models of dark energy. The philosophy is akin to what has been done in the case of single field quintessence models, where the analysis of exponential potentials and the classification of all fixed points allows for a qualitative understanding of more general cases. In the last section though, we will also explore a particular situation where both $k_1$ and $k_2$ exhibit  non-trivial functional dependence. 

It is evident, both from the equations of motion and from Fig. \ref{fig:phaseDiag},  that the system is symmetric under $\phi_1\rightarrow -\phi_1$, $k_1\rightarrow -k_1$ and $k_2\rightarrow -k_2$, and therefore one can focus on the $k_2>0$ half-plane without loss of generality. Given these assumptions one finds that $x_1>0$ as $\phi$ rolls down its potential. Through Eq. \eqref{eq:kprime} this implies that 
\begin{itemize}
\item $k_1$ grows when $1-\Gamma_1>0$ ,
\item $k_1$ decreases when $1-\Gamma_1<0$ .
\end{itemize}
We will therefore organise our analysis by considering these two regimes separately. In order to be systematic,  for each of these regimes, we will consider the subcases of steep ($k_2>\sqrt{3}$) and shallow ($k_2<\sqrt{3}$) potentials.

\section{Non-geodesic dynamics and the swampland bound}
\label{Sec3}

Let us now discuss an issue that serves as a motivation to perform this analysis and search for interesting alternative descriptions of dark energy in the context of effective theories with more than one scalar. As anticipated in the introduction, the dS swampland conjecture proposes an $\mathcal{O}(1)$ bound on the steepness of the scalar potential valid in those regions where the potential itself takes positive values. Such a bound would obviously rule out the existence of a dS vacuum, but in addition to this, it would also be in tension with any model of quintessence based on (effective) single-field dynamics.

On the other hand, a phase of accelerated expansion, like the one observed today or of the type postulated by the inflationary paradigm in the early universe requires
\be
\epsilon_H\equiv-\frac{\dot{H}}{H^2} < 1 \ .
\ee
Whenever the energy density of the universe is dominated by minimally coupled (both to gravity and among themselves) scalar fields, this requirement generically translates into a bound on the flatness of the potential
\be
\epsilon_V\equiv\frac{M_{\mathrm{Pl}}^{2}}{2}\left(\frac{V_\phi}{V}\right)^2\approx \epsilon_H<1\ ,
\ee
suggesting that flat potentials may be needed in order to support a phase of accelerated expansion.

While these flatness conditions are easily accommodated at the level of model building, it has been conjectured that consistent theories of quantum gravity might impose a lower bound on the flatness of the scalar potentials of the form \cite{Obied:2018sgi}
\be
\frac{V_\phi}{V} \ge \frac{c}{M_{\mathrm{Pl}}}\,,
\label{SwBound}
\ee
where $c$ is an $\mathcal{O}(1)$ constant parameter, that can in principle be derived from the UV theory. Such a lower bound is in clear tension with the desire for scalar field driven accelerated cosmic expansion, both at early and at late times.

In fact it has been shown in \cite{Agrawal:2018own} that in the context of quintessence models with exponential potentials of the form $V=V_0 \,e^{-k_2 \phi/ M_{Pl}}$, only models with $c=k_2<0.6$ remain compatible with the latest data \cite{Scolnic:2017caz}, thus highlighting the tension between the swampland bound and cosmological observations. As stressed in the Sec. \ref{Intro}, these models, besides lacking an explicit string embedding, would also share the same control issues of dS vacua with additional phenomenological constraints coming from fifth-forces.

It is worth noting that the current observational bounds constrain not only the present $\omega_\phi$ and $\Omega_\phi$ but also the time variation of $\omega_\phi$ at low redshifts. For the purposes of this paper we neglect the redshift dependence of $\omega_\phi$ and consider the observational window to be given by $\Omega_\phi = 0.7$ and $\omega_\phi=-1$ with an uncertainty of $\pm 0.09$ at  $95\% $ CL in the equation of state parameter, which we represent by two red dots in the $(\omega_\phi,\Omega_\phi)$ plane.\footnote{We consider these numbers as a ballpark measure of the dark energy parameters, specific dataset combinations yield slightly different figures that lie in this range, see \cite{Scolnic:2017caz}.}

One way to alleviate this tension is to consider kinetically coupled scalar fields (through non-canonical kinetic terms) and work in the regime of large field space curvature. This way, one gains the opportunity to support a small $\epsilon_H$ by using scalar potentials which are not at all flat, provided though that the field space trajectories selected by the dynamics of the system by suitable initial conditions have a sharp turning rate along non-geodesic trajectories.\footnote{Notice that in the multifield case the bound (\ref{SwBound}) generalizes to $\sqrt{\gamma^{ij} V_i V_j}/V\ge c/M_{\mathrm{Pl}}$ which would however still reduce to $V_{\phi_1}/V\ge c/M_{\mathrm{Pl}}$ for the case under consideration described by the action (\ref{eq:S}).}

It is perhaps worth mentioning that such constructions, which could in principle yield an arbitrarily high number of efoldings of cosmic acceleration, would not in principle be in manifest contrast with the distance swampland conjecture. This is because it has been argued that the conjecture would only pose sharp constraints on purely geodesic distances, and these would not grow very large in these models. 
More quantitatively, the geodesic distance can be defined as 
\be
\Delta \phi\equiv \int dt \sqrt{\gamma_{ij} \dot{\phi}^i\dot{\phi}^j}\ .
\ee
Considering the field space metric as implicitly defined in Eq. \eqref{eq:S} and putting the fields on-shell (assuming negligible accelerations) it can be shown that 
\be
\frac{\Delta \phi}{M_{Pl}}= \int dN \sqrt{-\left(\frac{3}{2k_1}\right)^2+3\,\frac{k_2}{k_1}}\,,
\ee
where $N\equiv \ln a$ is the number of efoldings of expansion. In the regime $k_1\gg k_2\sim\mathcal{O}(1)$ and noting that for quintessential dynamics $\Delta N\le10$ it seems natural to conclude that parametrically
\be
\frac{\Delta \phi}{M_{\mathrm{Pl}}}\sim \frac{\Delta N }{\sqrt{k_1}} <1\ ,
\ee 
in accordance with swampland distance bounds.

\section{Growing kinetic coupling}
\label{Sec4}

As anticipated earlier, our systematic analysis starts from the case in which the kinetic coupling $k_1$ is a growing function of $\phi_1$.
In this situation, we had seen that the inequality $1-\Gamma_1>0$ holds. This regime encompasses several interesting kinetic couplings:\footnote{From this point on, in order to reduce clutter in the formulae, we set $M_\mathrm{Pl}=1$.}
\begin{itemize}
\item  $f=\ln(\phi_1^p)$, implying $k_1=\frac{-1}{\phi_1 \ln\phi_1}$ and  $1-\Gamma_1=1+\ln\phi_1$ with $\phi_1>1/e$;
\item $f=\phi_1^p$, implying $k_1=-\frac{p}{\phi_1}$ and $1-\Gamma_1=\frac{1}{p}$ with $p>0$;
\item $f=e^{\alpha \phi_1^p}$, implying $k_1=-p \alpha\phi_1^{p-1}$ and $1-\Gamma_1=\frac{(1-p)}{p \alpha}\phi_1^{-p}$ 
in the regime $\alpha>0$ and $p\in[0,1]$ or alternatively $\alpha<0$ and ($p<0$ or $p>1$).
\end{itemize}
In all these cases the system will tend to evolve towards a $\mathcal{NG}$ fixed point in the asymptotic future. The dynamics of the transition from a matter dominated regime to $\mathcal{NG}$ will differ depending on the steepness of the scalar potential, on the particular form of the function $f(\phi_1)$, and on the initial conditions chosen for the vacuum expectation value of $\phi_1$. 

\begin{figure}[h!]
	\centering
	\begin{minipage}[b]{0.31\linewidth}
	\centering
	\includegraphics[width=\textwidth]{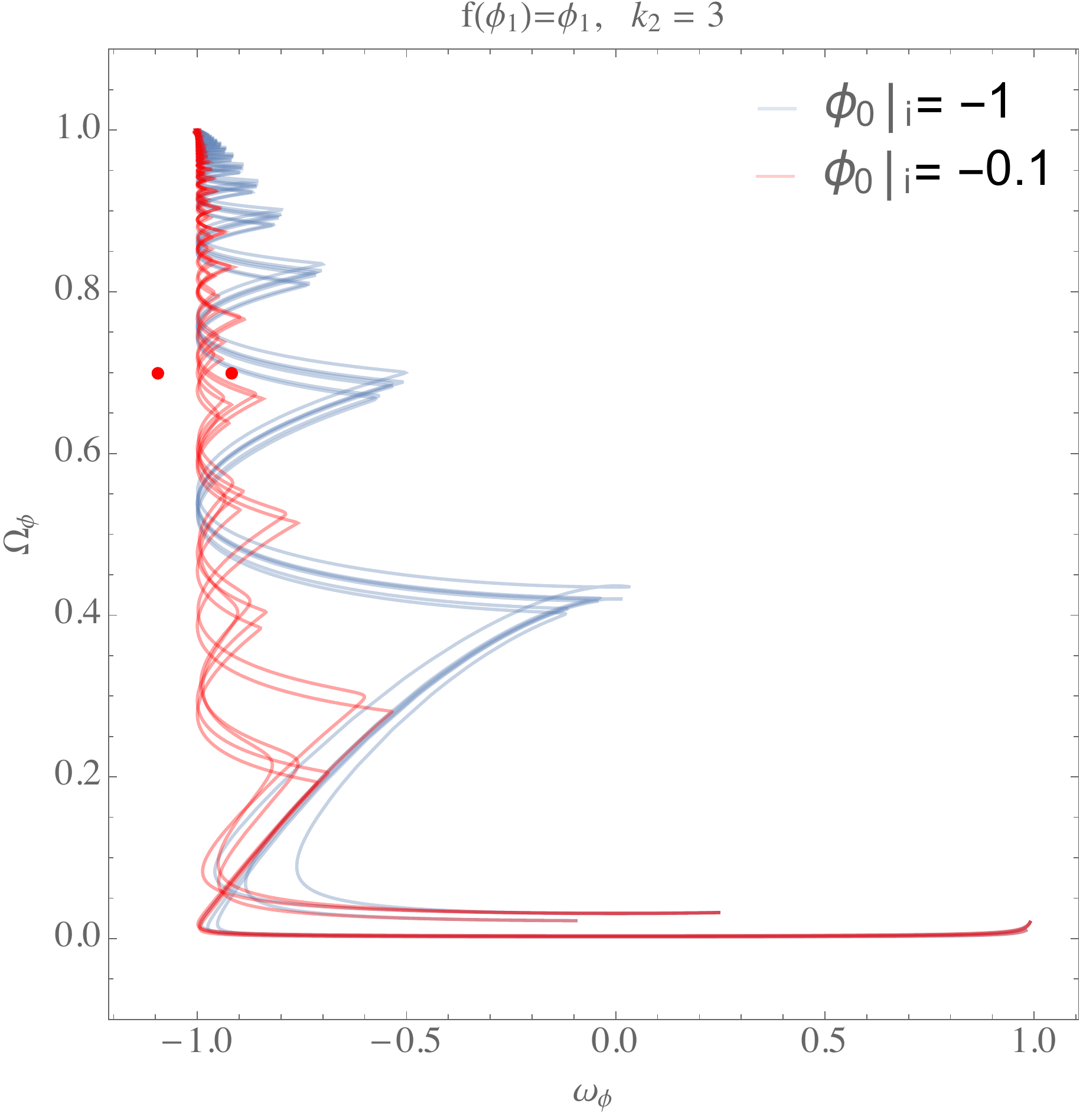}
    \end{minipage}
	\hspace{0.2cm}
	\begin{minipage}[b]{0.31\linewidth}
	\centering
	\includegraphics[width=\textwidth]{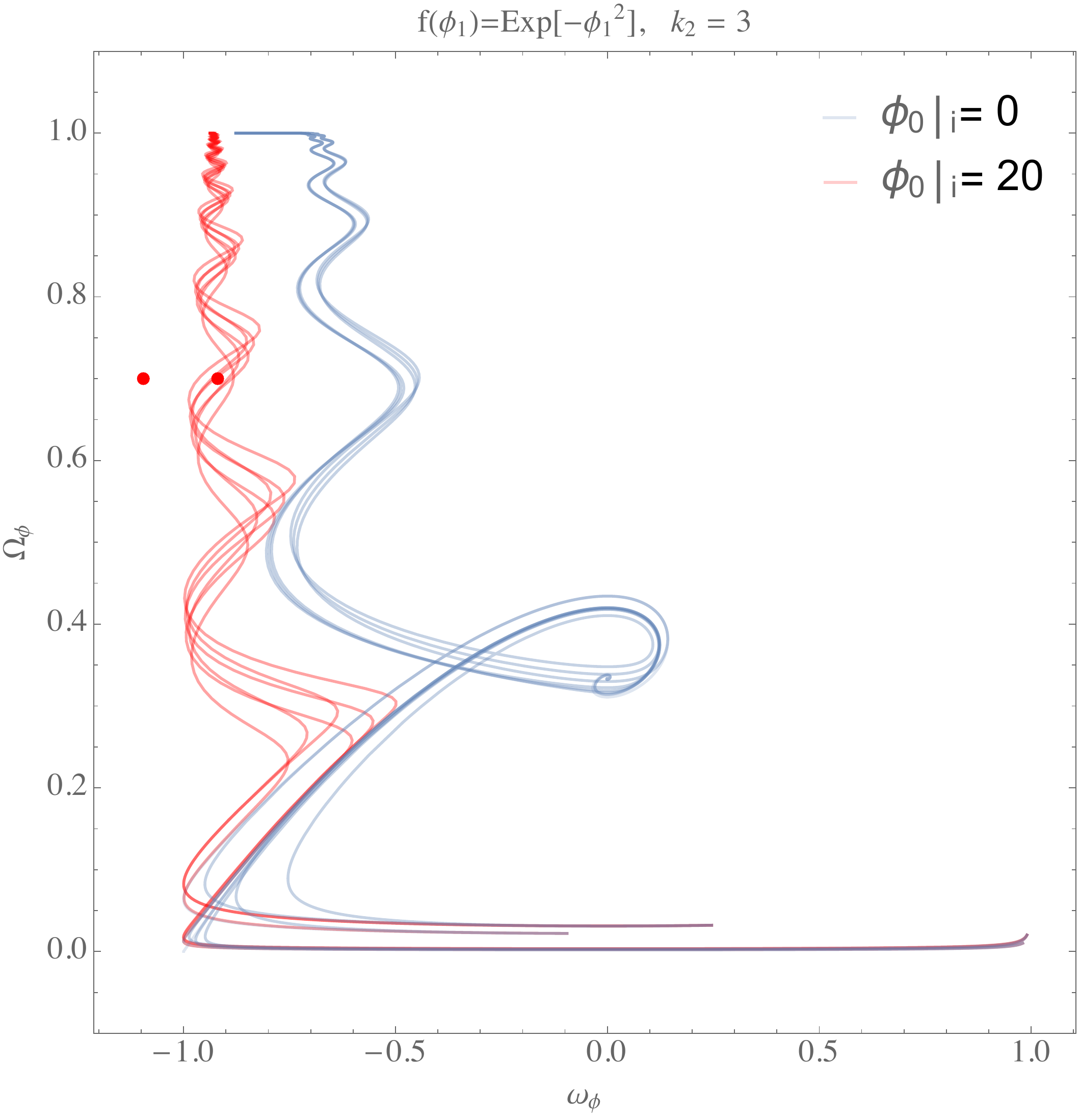}
	\end{minipage}
	\hspace{0.2cm}
	\begin{minipage}[b]{0.31\linewidth}
	\centering
	\includegraphics[width=\textwidth]{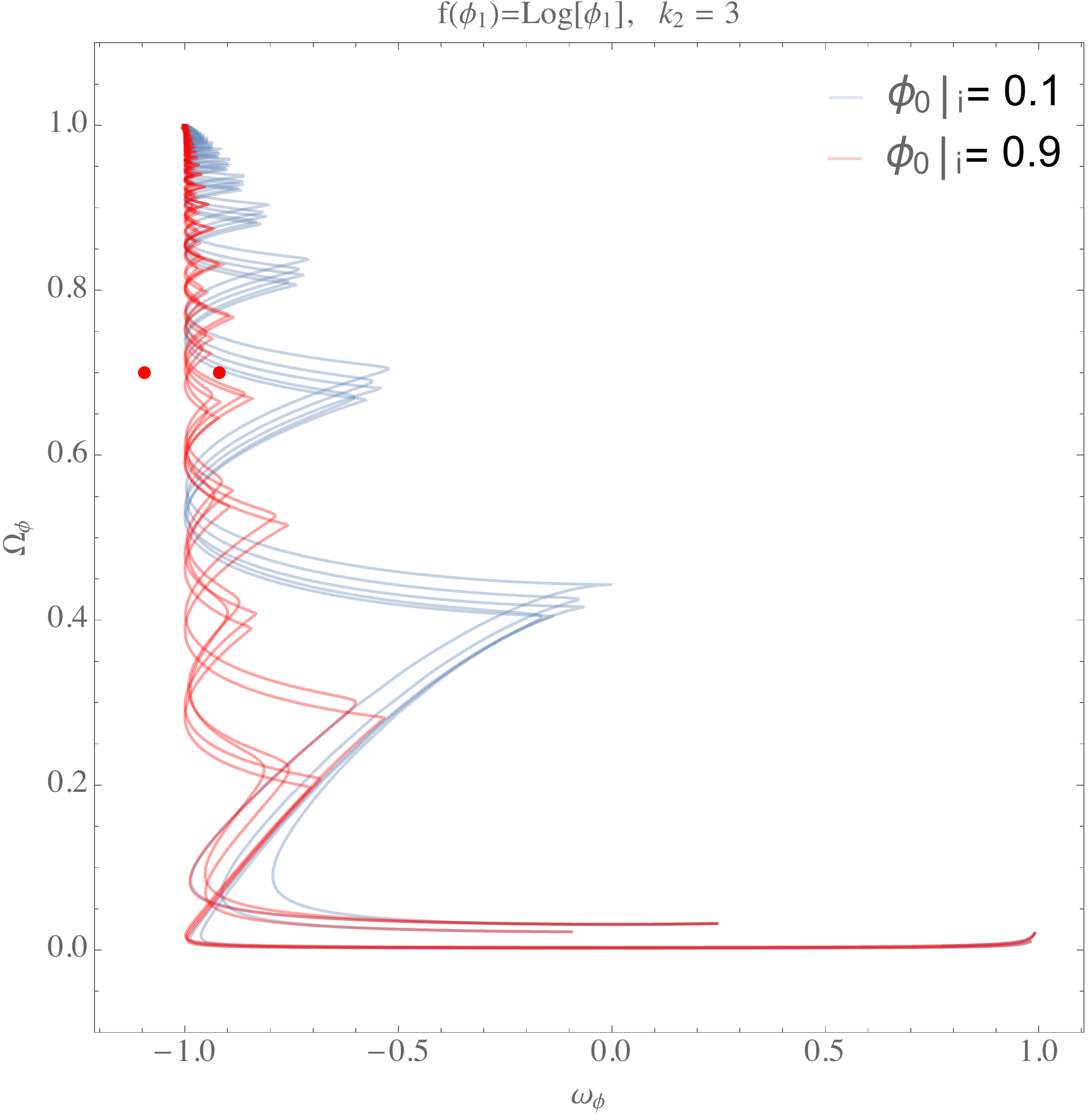}
	\end{minipage}
	  \caption{($\Omega_\phi, \omega_\phi)$ plane evolution for matter dominated initial conditions and $k_2=3$. Trajectories that seem to end at $\mathcal{S}$ without passing through $\mathcal{NG}$ should be considered as a numerical artifact.}
\label{fig:growingSteep}
\end{figure}

\paragraph{Steep scalar potentials:} for steep scalar potentials, with $k_2>\sqrt{3}$, if the initial conditions for $\phi_1$ are such that $\phi_1|_{i}\ll \phi_{1\times} $, where
\be
k_1(\phi_{1\times}) = \frac{k_2}{2} \ ,
\label{eq:k1crossSteep}
\ee
the system will initially evolve towards the scaling fixed point $\mathcal{S}$, before settling into the $\mathcal{NG}$ regime. Several examples are illustrated in Fig. \ref{fig:growingSteep}, where the blue trajectories correspond to initial conditions that lead to a transition $\mathcal{S}\rightarrow\mathcal{NG}$, whereas the red trajectories converge directly to $\mathcal{NG}$. We see that the marked oscillations in the transition between the scaling and the $\mathcal{NG}$ regimes make it difficult to obey observational bounds on $\omega_\phi$ (given by line between the two red dots). Initial conditions that lead directly to $\mathcal{NG}$ are in that respect more promising, despite also featuring a strong oscillatory behavior that inevitably leads to a tension with the bounds on the variation of $\omega_{\phi}$ at low redshifts. 

In what concerns the initial evolution and the final state of the system, this is qualitatively similar to the scaling-freezing models in single field quintessence where the potential takes the form
\be
V(\phi)=e^{-k_2|_i \phi}+e^{-k_2|_f \phi}\ ,
\ee
where $k_2|_i >\sqrt{3}$ and  $k_2|_f <\sqrt{3}$, even though in our case the steepness of the potential is kept fixed. A crucial difference with scaling-freezing models arises from the fact that the oscillatory motion in the transient between $\mathcal{S}$ and $\mathcal{NG}$ makes it difficult to account for the current state of the universe with this type of dynamics.

\begin{figure}[h!]
	\centering
	\begin{minipage}[b]{0.31\linewidth}
	\centering
	\includegraphics[width=\textwidth]{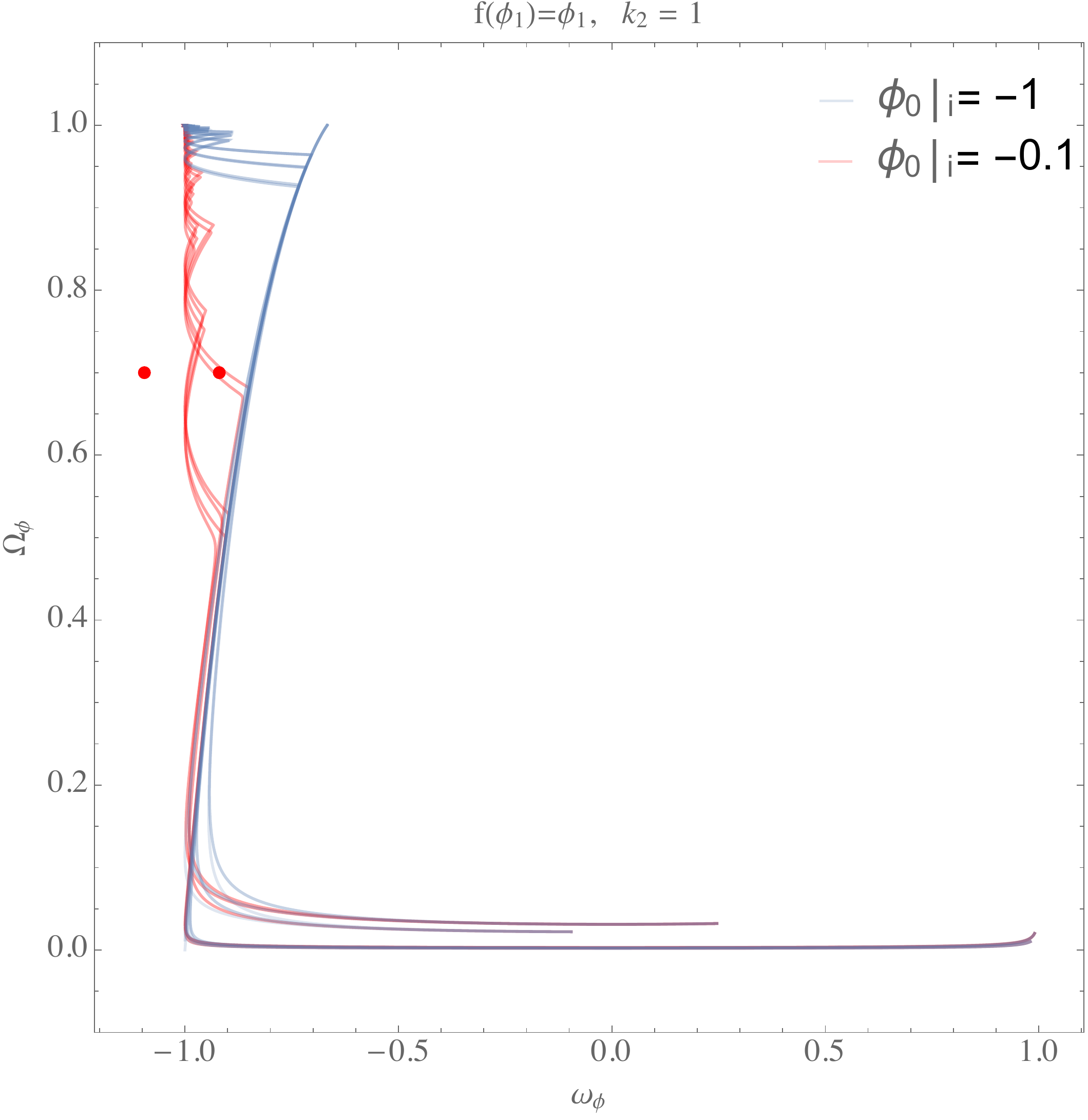}
    \end{minipage}
	\hspace{0.2cm}
	\begin{minipage}[b]{0.31\linewidth}
	\centering
	\includegraphics[width=\textwidth]{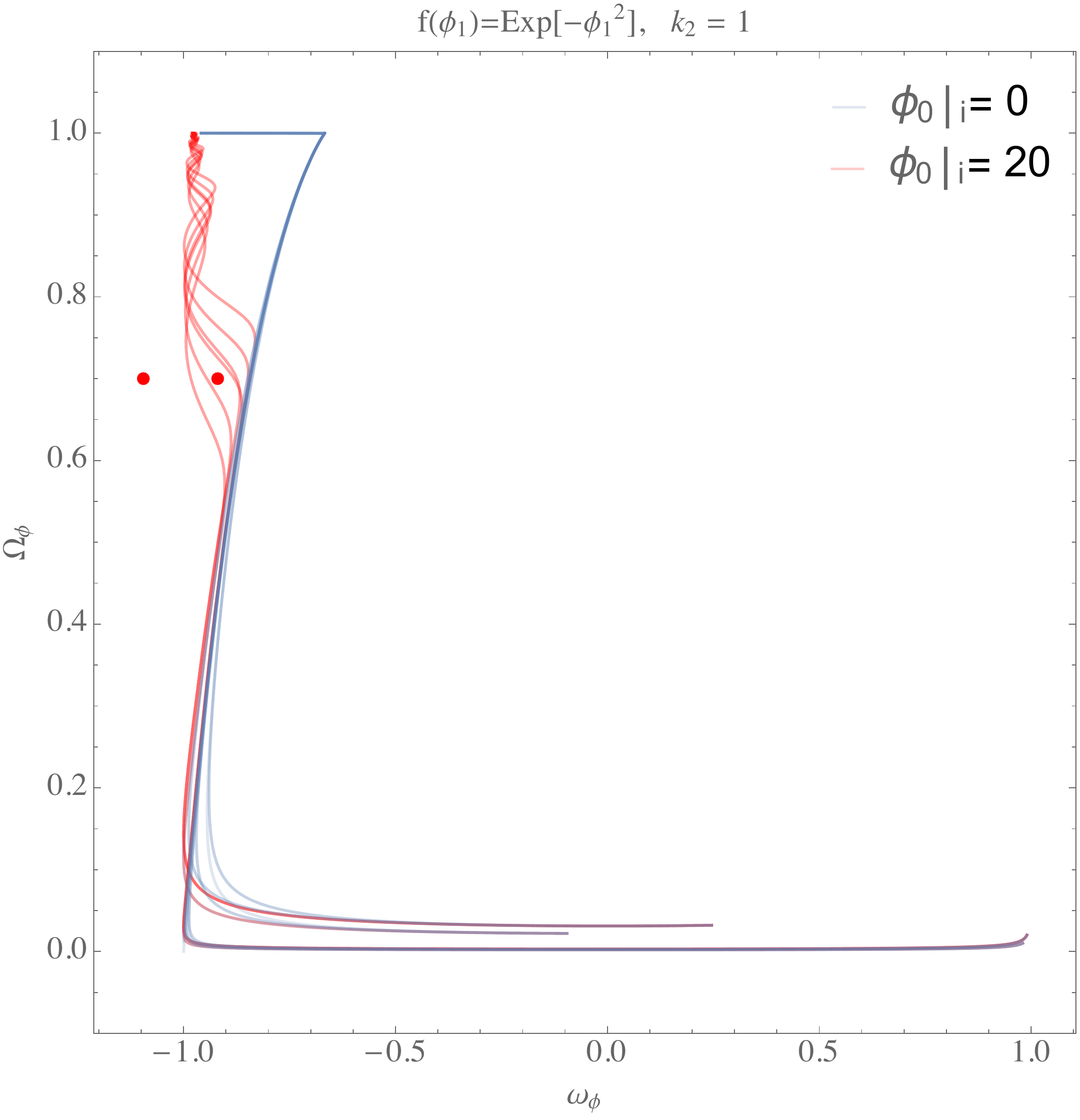}
	\end{minipage}
	\hspace{0.2cm}
	\begin{minipage}[b]{0.31\linewidth}
	\centering
	\includegraphics[width=\textwidth]{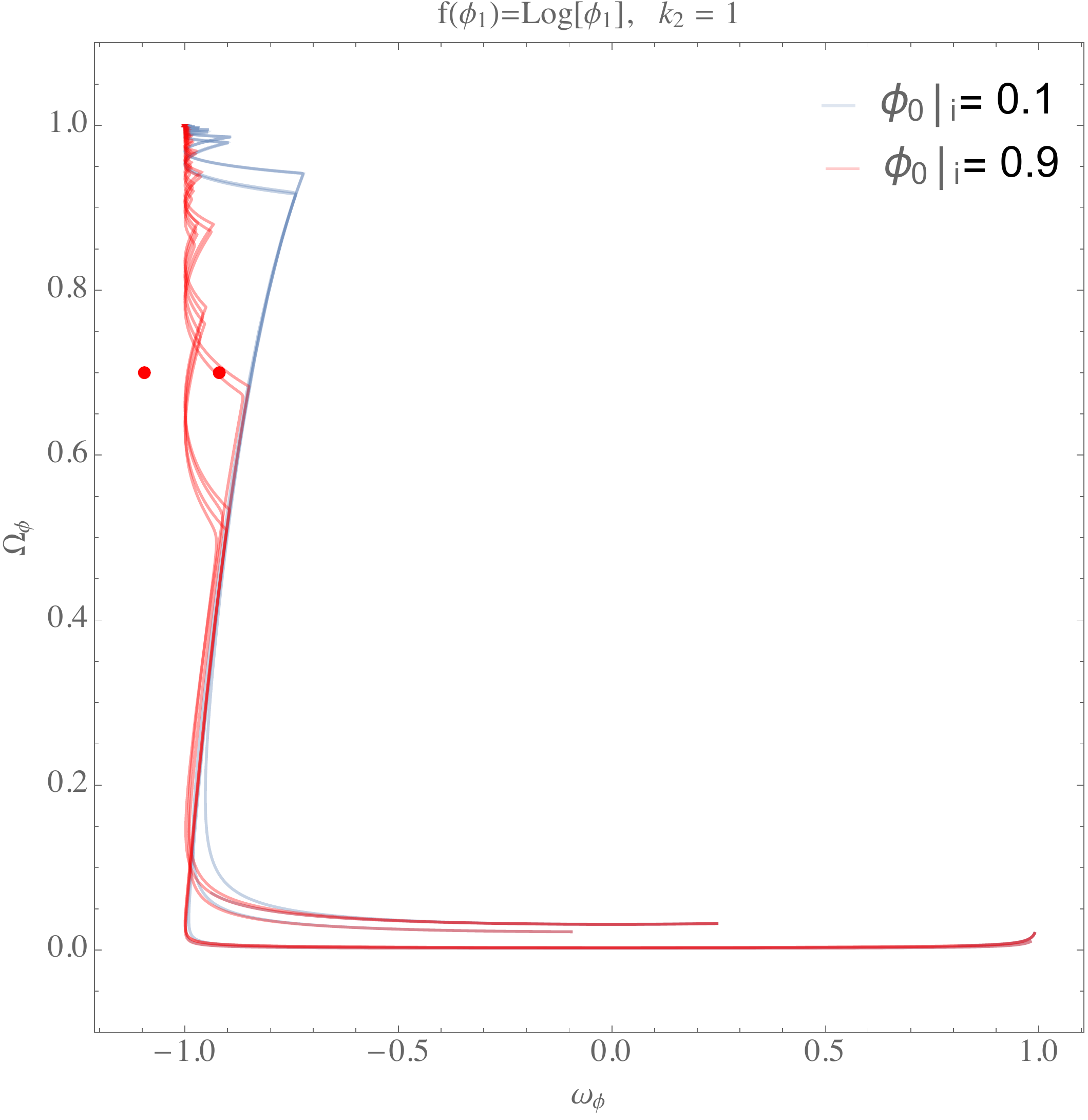}
	\end{minipage}
	  \caption{($\Omega_\phi, \omega_\phi)$ plane evolution for matter dominated initial conditions and $k_2=1$.}
\label{fig:growingShallow}
\end{figure}

\paragraph{Shallow scalar potentials:} if the potential is too shallow to support a scaling regime, $k_2<\sqrt{3}$, the system can initially converge into a scalar dominated $\mathcal{G}$
 fixed point, before eventually settling into the $\mathcal{NG}$ regime. This happens provided  $\phi_1|_{i}\ll \phi_{1\times} $, where now 
\be
k_1(\phi_{1\times} )= \frac{6-k_2^2}{2 k_2}\ .
\label{eq:k1crossShallow}
\ee
Should the initial conditions parametrically violate this requirement, the system will then evolve directly towards $\mathcal{NG}$.
An example is illustrated in Fig. \ref{fig:growingShallow}. We observe that viable transients that are compatible with the swampland bound $k_2\ge1$ are more easily attained if the initial conditions cause the system to evolve direclty towards $\mathcal{NG}$.

\section{Decaying kinetic coupling}
\label{Sec5}

\begin{figure}[h!]
	\centering
	\begin{minipage}[b]{0.31\linewidth}
	\centering
	\includegraphics[width=\textwidth]{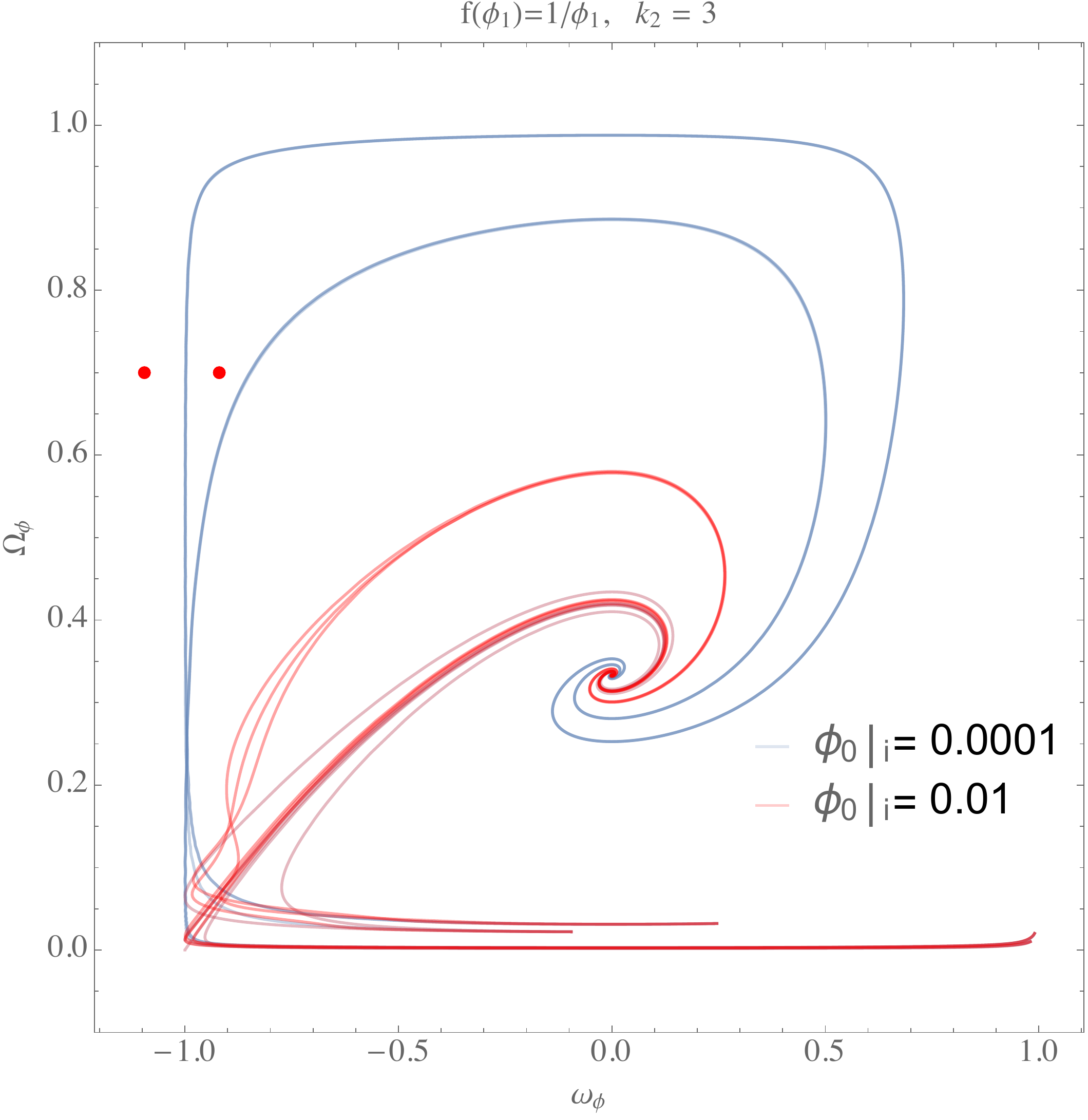}
    \end{minipage}
	\hspace{0.2cm}
	\begin{minipage}[b]{0.31\linewidth}
	\centering
	\includegraphics[width=\textwidth]{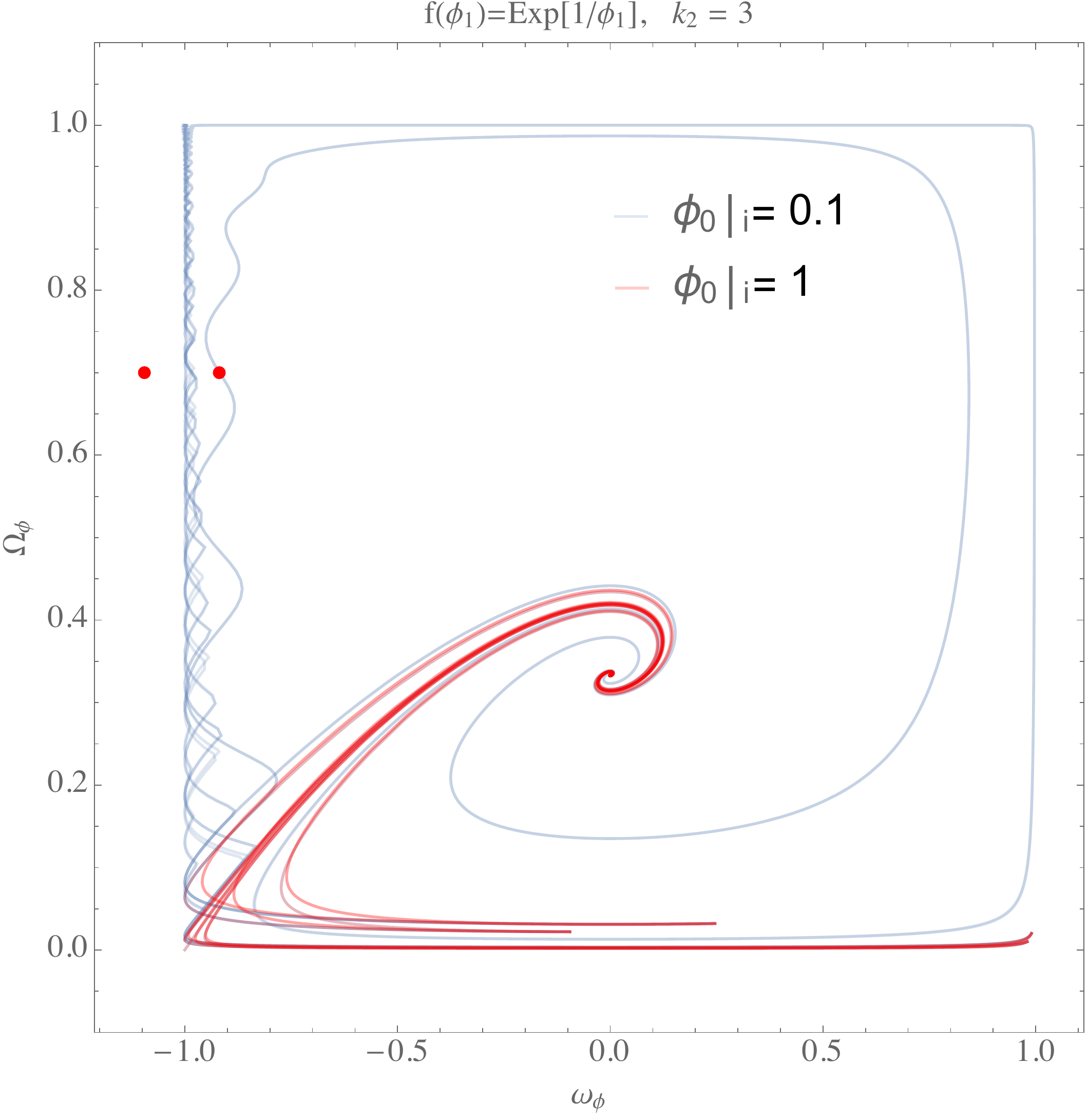}
	\end{minipage}
	\hspace{0.2cm}
	\begin{minipage}[b]{0.31\linewidth}
	\centering
	\includegraphics[width=\textwidth]{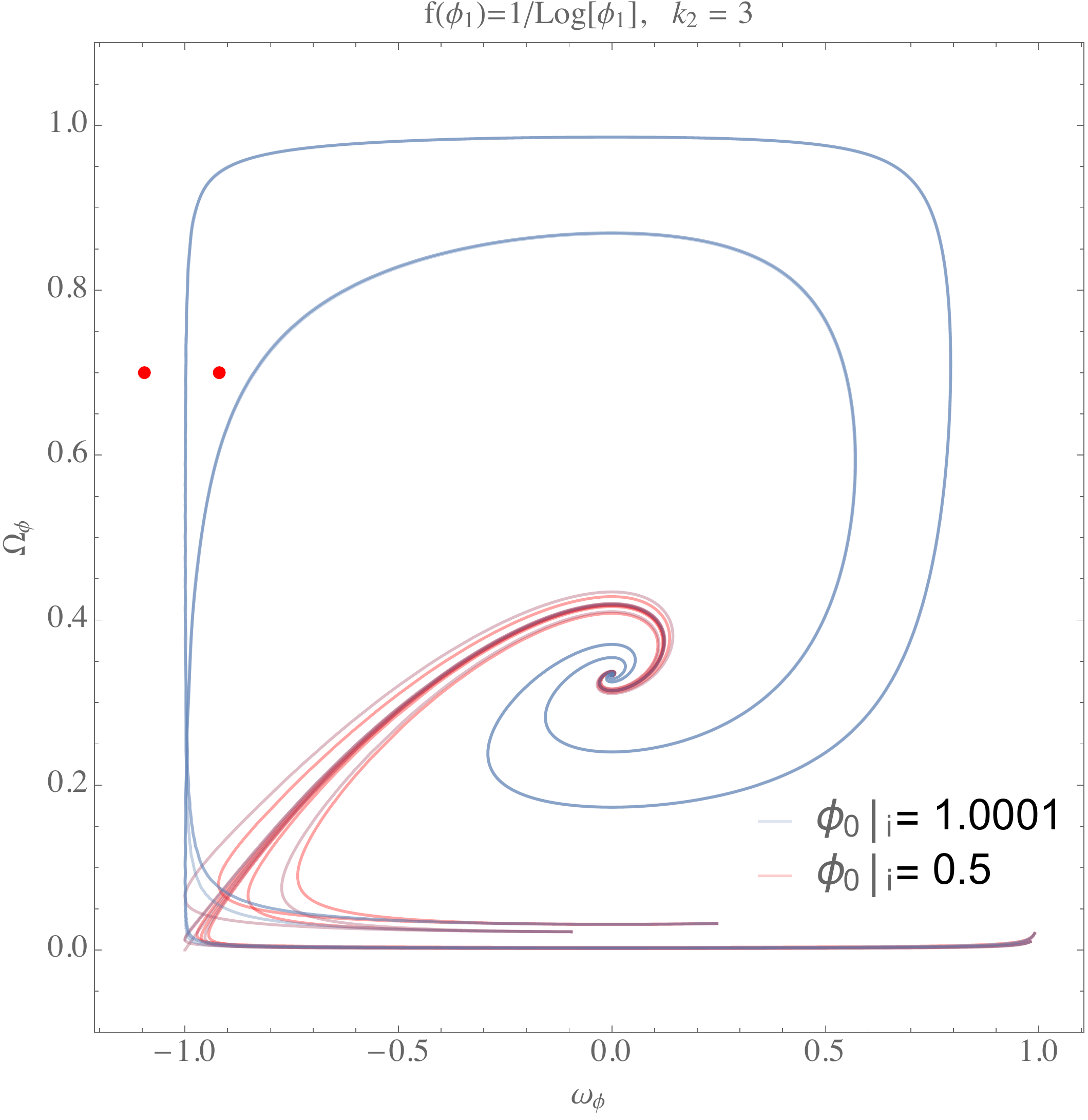}
	\end{minipage}
	  \caption{($\Omega_\phi, \omega_\phi)$ plane evolution for matter dominated initial conditions and $k_2=3$.}
\label{fig:decayingSteep}
\end{figure}

In this regime the kinetic coupling is such that $k_1$ decreases as $\phi_1$ rolls down its potential. These trajectories will converge at late times either towards a scaling or a geodesic fixed point, depending on the steepness of the scalar potential. Just like in Sec. \ref{Sec4} there are various kinetic couplings that can lead to this behavior:
\begin{itemize}
\item  $f=1/\ln(\phi_1^p)$, implying  $k_1=\frac{p}{\phi_1\ln \phi_1}$ and $1-\Gamma_1=-1-\ln\phi_1$ with $\phi_1>1/e$;
\item $f=\phi_1^p$, implying $k_1=-\frac{p}{\phi_1}$ and $1-\Gamma_1=\frac{1}{p}$ with $p<0$;
\item $f=e^{\alpha \phi_1^p}$, implying $k_1=-p \alpha \phi_1^{p-1}$ and $1-\Gamma_1=\frac{(1-p)}{p \alpha}\phi_1^{-p}$ 
in the regime $\alpha<0$ and $p\in[0,1]$ or alternatively $\alpha>0$ and ($p<0$ or $p>1$).
\end{itemize}
Once more we can assume that $k_2>0$ and consider the $k_2>\sqrt{3}$ and $k_2<\sqrt{3}$ separately, i.e. distinguishing between steep and shallow potentials, respectively. 
\paragraph{Steep scalar potentials:} for steep scalar potentials, the border between the stability regions of $\mathcal{S}$ and $\mathcal{NG}$ is once again given by Eq. \eqref{eq:k1crossSteep}.
Initial field displacements  $\phi_1|_{i}\ll \phi_{1\times}$ lead to an initial transient of large $k_1$, which pushes the system towards the $\mathcal{NG}$ fixed point. As $\phi_1$ increases $k_1$ decreases and the system will eventually settle into the $\mathcal{S}$ fixed point, as illustrated in Fig. \ref{fig:decayingSteep}. Initial conditions leading to a smaller initial value of $k_1$ converge directly towards the late time attractor $\mathcal{S}$.

\begin{figure}[h!]
	\centering
	\begin{minipage}[b]{0.31\linewidth}
	\centering
	\includegraphics[width=\textwidth]{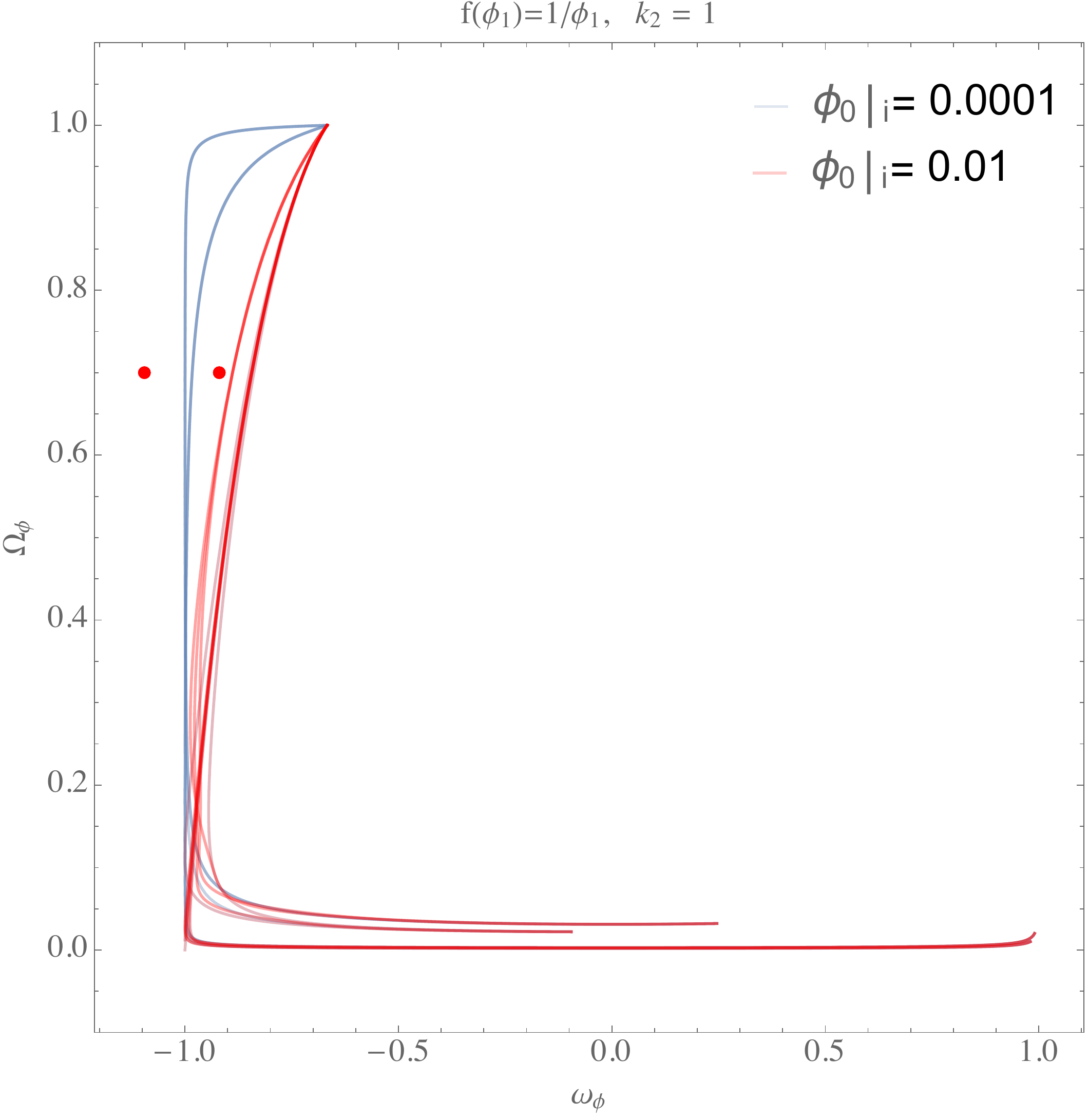}
    \end{minipage}
	\hspace{0.2cm}
	\begin{minipage}[b]{0.31\linewidth}
	\centering
	\includegraphics[width=\textwidth]{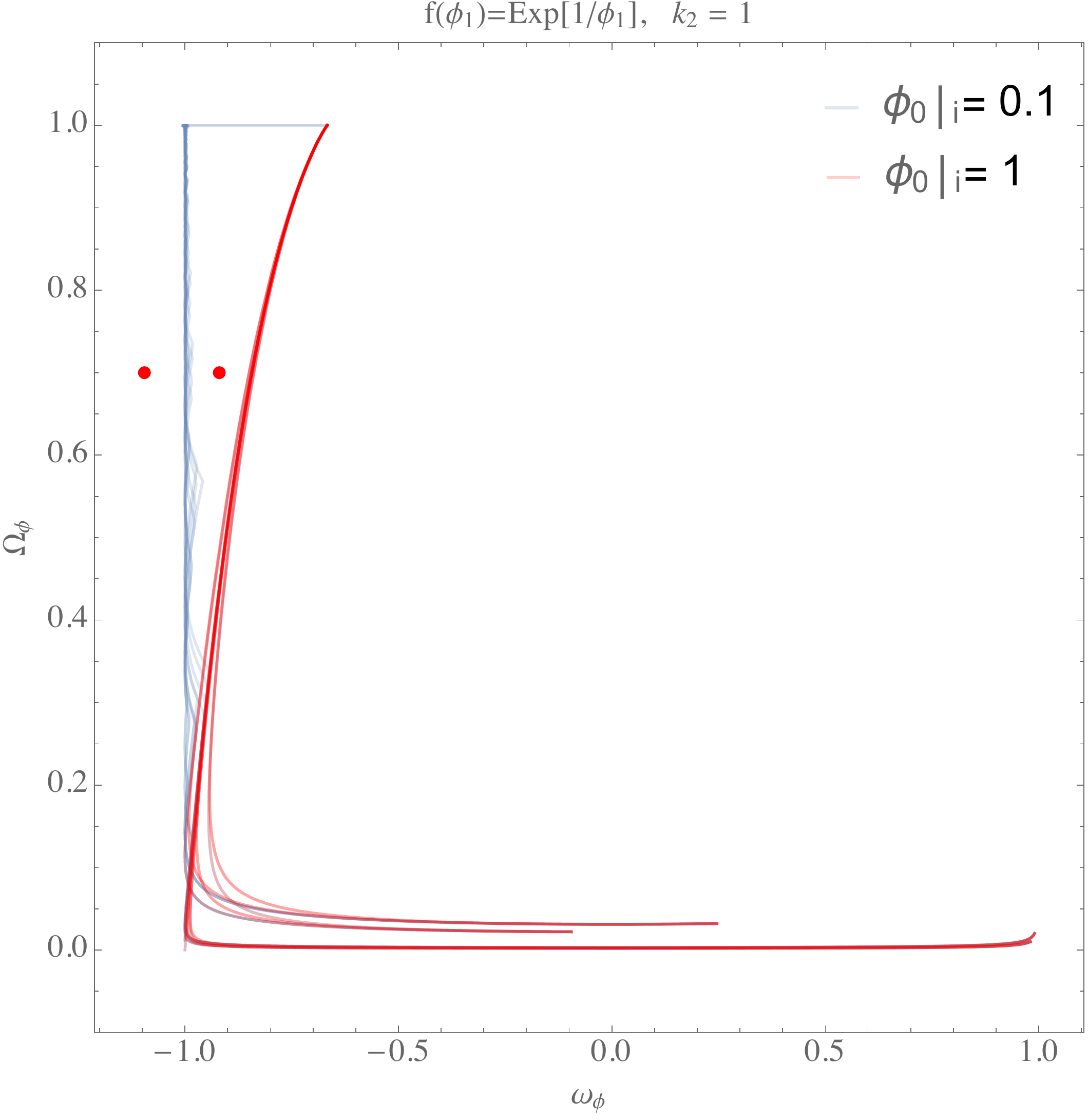}
	\end{minipage}
	\hspace{0.2cm}
	\begin{minipage}[b]{0.31\linewidth}
	\centering
	\includegraphics[width=\textwidth]{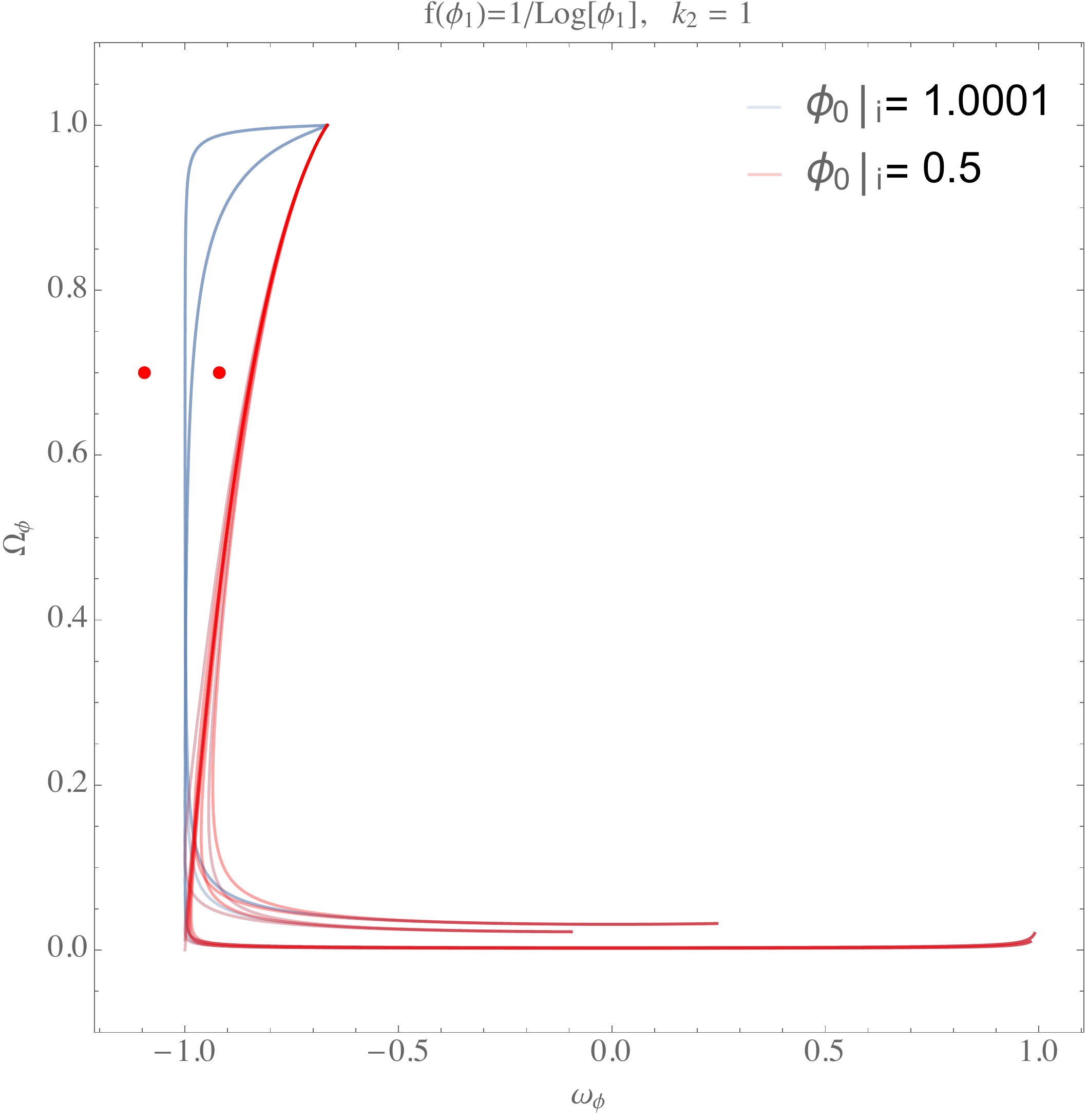}
	\end{minipage}
	  \caption{($\Omega_\phi, \omega_\phi)$ plane evolution for matter dominated initial conditions and $k_2=1$.}
\label{fig:decayingShallow}
\end{figure}

\paragraph{Shallow scalar potentials:} the border between the stability domains of $\mathcal{G}$ and $\mathcal{NG}$ is given by the solution of \eqref{eq:k1crossShallow} for each kinetic coupling, and therefore if  $\phi_1|_{i}\ll \phi_{1\times}$ the system will initially converse to $\mathcal{NG}$ before settling into $\mathcal{G}$, as illustrated in Fig. \ref{fig:decayingShallow}. This type of dynamics allows for observational viable transients even in the presence of steep potentials.

\section{Non-exponential potentials}
\label{Sec6}

The analysis of Sec. \ref{Sec4} and \ref{Sec5} and of \cite{Cicoli:2020cfj} relied, at least partially, on the simplifications afforded by the assumption of exponential potentials and/or kinetic functions. In order to better understand the dynamics of the system in a more general manner, in this section we extend our analysis by simultaneously considering both non-exponential potentials and kinetic couplings. Given the wide space of possible functional combinations for $V$ and $f$ and our reliance on numerical methods, we will focus on the simplest cases, where both potential and kinetic coupling are monomial functions of $\phi_1$:
\be
V=V_0\  \phi_1^{p_2}\qquad\text{and}\qquad f=f_0\ \phi_1^{p_1}\,.
\ee
For this choice we find 
\be
k_1= -\frac{p_1}{\phi_1}\qquad\text{and}\qquad k_2= -\frac{p_2}{\phi_1}\,.
\ee
In this special case, though $k_1$ and $k_2$ are field dependent, their ratio is not
\be 
\frac{k_1}{k_2}=\frac{p_1}{p_2}\ .
\ee
One therefore sees that the trajectories in the $(k_1,k_2)$ plane are straight lines through the origin. The direction of the trajectories depends on the sign of $p_2$: for $p_2<0$, $\phi_1$ grows with the expansion and so at late times $k_1,k_2\rightarrow 0$. Conversely, for $p_2>0$, $\phi_1$ decreases with the expansion causing $k_1,k_2\rightarrow \infty$. Clearly $\mathcal{NG}$ dynamics can play a role in the expansion of the universe only if $k_1$ and $k_2$, or equivalently $p_1$ and $p_2$, have the same sign (see Fig \ref{fig:phaseDiag}), and hence we will focus exclusively on these cases.

The equation-of-state parameter for the $\mathcal{NG}$ fixed point is constant, even though $k_1$ and $k_2$ are not, and fixed by the ratio $p_1/p_2$ as
\be
\omega_\mathcal{NG}=\frac{p_2-2 p_1}{2p_1+p_2}\ ,
\ee
and thus viable trajectories can be found in the regime where $p_1$ and $p_2$ have the same sign and $|p_1 |\gg |p_2|$.

\begin{figure}[h!]
	\centering
	\begin{minipage}[b]{0.45\linewidth}
	\centering
	\includegraphics[width=\textwidth]{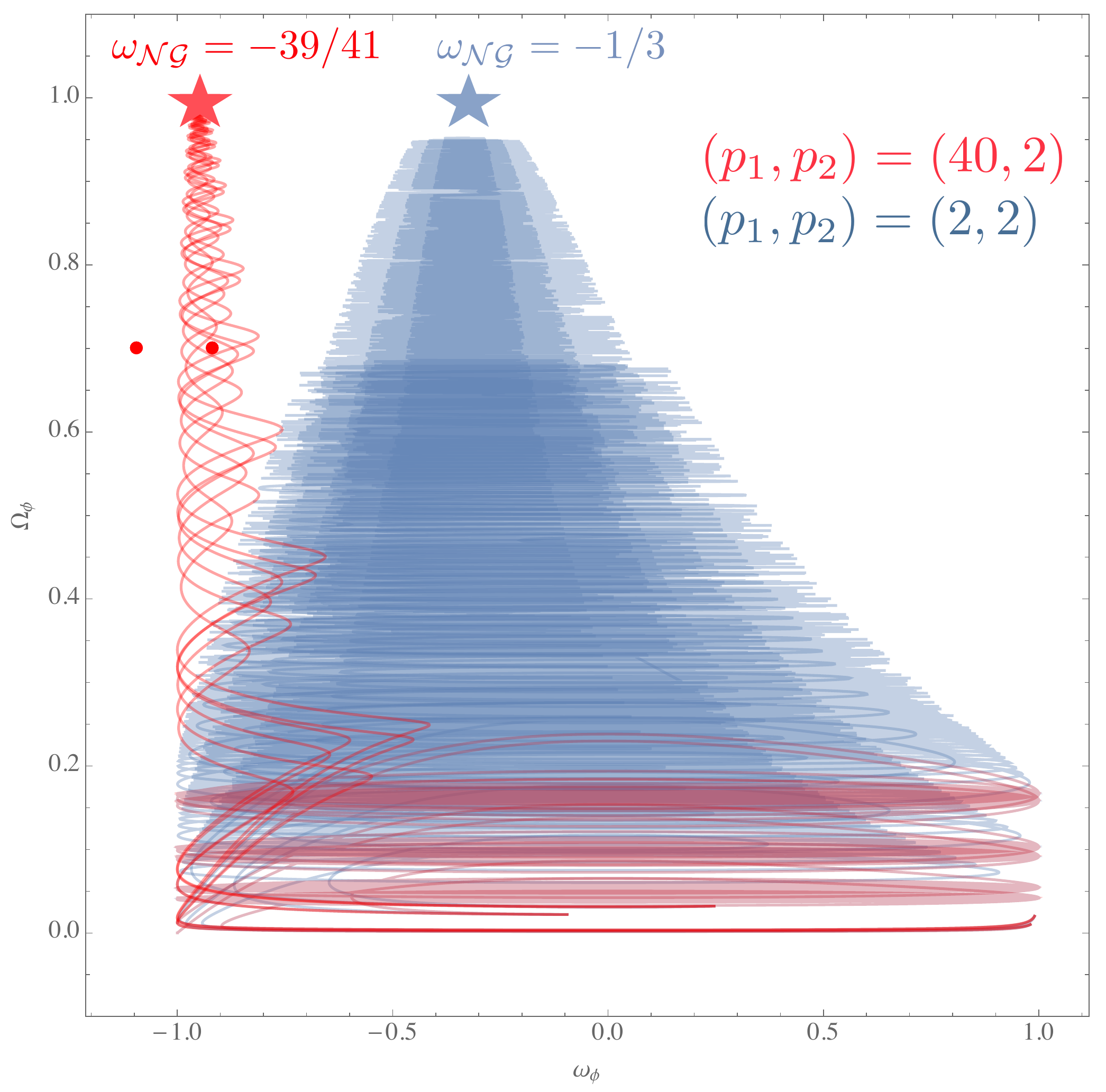}
    \end{minipage}
	\hspace{0.2cm}
	\begin{minipage}[b]{0.45\linewidth}
	\centering
	\includegraphics[width=\textwidth]{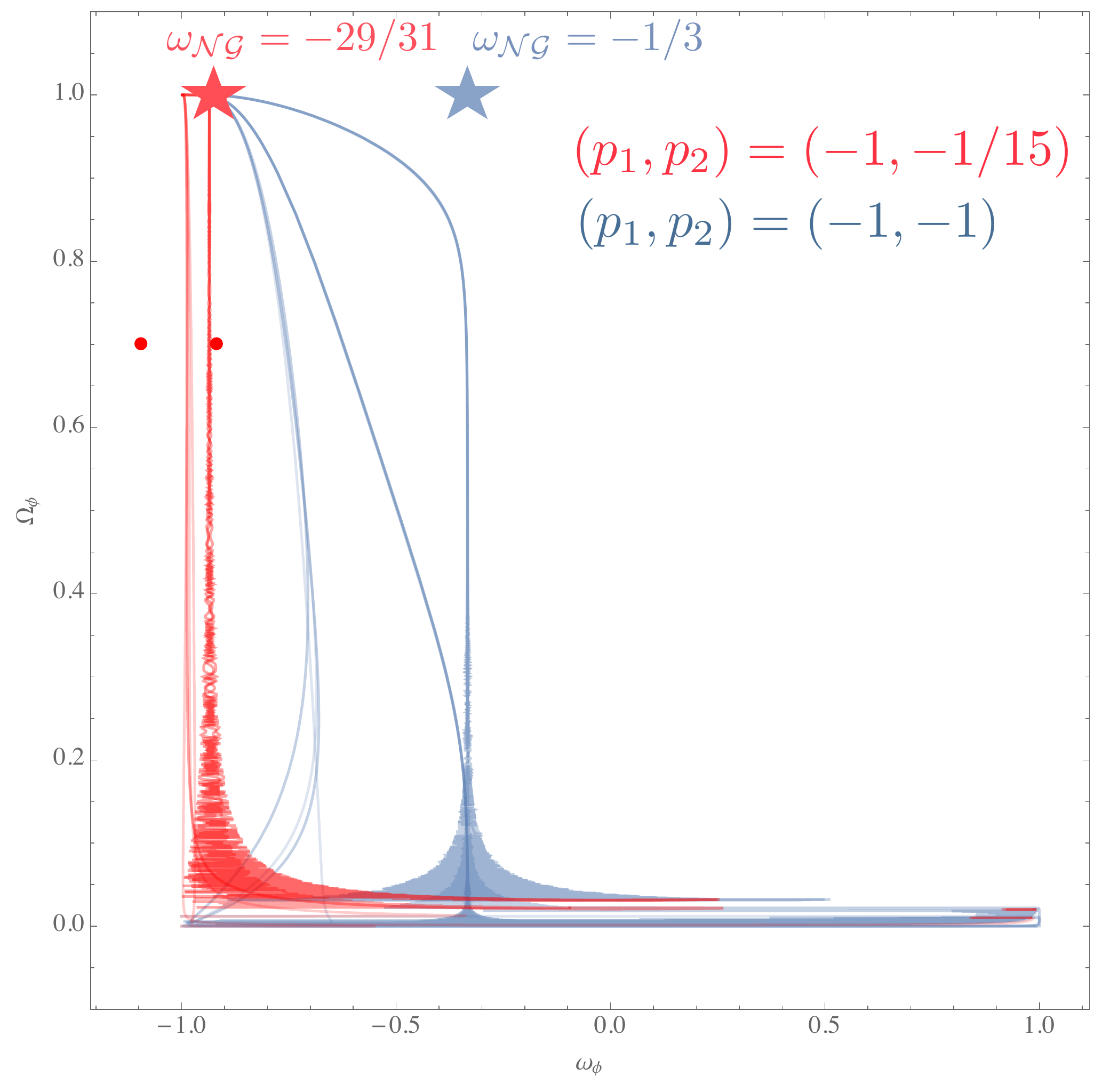}
	\end{minipage}
	  \caption{($\Omega_\phi, \omega_\phi)$ plane evolution for matter dominated initial conditions and $k_1= -\frac{p_1}{\phi_1}$ and $ k_2= -\frac{p_2}{\phi_1}$.}
\label{fig:p1p2}
\end{figure}

In Fig. \ref{fig:p1p2} we plot two representative examples of the $k_1,k_2>0$ and $k_1,k_2<0$ regimes. We see that an appropriate choice of initial conditions allows the system to converge to $\mathcal{NG}$ irrespective of the signs of $k_1, k_2$, provided they are the same. For the case $k_1,k_2>0$ one chooses $\phi_1|_i$ small ($10^{-5}$ in the figure) in order to have $k_1,k_2\gg1$.

For $k_1,k_2<0$, $\phi_1$ decreases with expansion and the system approaches the regime $k_1,k_2\ll-1$. Note that in this case, in order to avoid slow-roll/$\mathcal{G}$ dynamics one must have $\phi_i|_{\mathrm{initial}}<\sqrt{p_2}$ (in the figure we set it to $1/2$).

Following this criterion for the choice of initial conditions, one finds that though the trajectories followed by the system approach the instantaneous fixed point $\mathcal{NG}$, they do so in a significantly different manner: for $k_1,k_2<0$ the motion is dominated by the oscillations of the field around the minimum of its potential, leading to oscillatory trajectories in the $(\omega_\phi,\Omega_\phi)$ plane, which make it difficult to obey current constraints on the equation of state for dark energy, regardless of the relative size of $p_1$ and $p_2$. On the other hand for $k_1,k_2>0$ such oscillations are absent (the potential has no local minimum) and provided $|p_1|\gg |p_2|$ one can find trajectories with viable transients where the dynamics is inherently multi-field.

\section{Conclusions and Outlook}
\label{Sec7}

In this paper we have studied the dynamics of Einstein gravity coupled to a system of two scalar fields with a curved field space with a rather general functional dependence of the kinetic coupling function and the scalar potential. Our aim has been to find viable quintessence models that could effectively describe a phase of accelerated expansion like the one which is observed in late-time cosmology even for steep potentials, generalizing our earlier work \cite{Cicoli:2020cfj}. Interestingly, as opposed to the single field case, this multi-field situation can account for a quintessence phase without being manifestly in tension with possible lower bounds on the steepness of the corresponding scalar potential arising from the dS swampland conjecture. The richer structure present here is due to the existence of a novel fixed point in the phase diagram of the system, which crucially relies on genuinely multifield effects such as non-geodesic motion in field space.

In particular, we analyzed a situation where a non-canonical kinetic coupling with various functional dependences, together with a suitable choice of initial conditions on the field displacements, can give rise to explicit realizations of the aforementioned phenomenon. When restricting ourselves to exponential scalar potentials with an exactly flat direction, our analysis turns out to be exhaustive. To conclude, we also exhibited explicit examples of models consistent with the current cosmological data, where the scalar potential is not of an exponential form. Though a wide variety of cases are still not covered by our present work, we already have evidence for a very interesting structure in the most general case.

It is certainly worth mentioning that, while the models studied here are motivated by the recent dS swampland conjecture and offer a new spectrum of possibilities for constructing a UV embedding of dark energy, it remains to be seen whether any of these appealing features survive once an honest and complete top-down construction is considered. Addressing this question in detail is of utmost importance and we hope to come back to it in the future. For the time being, let us just mention that the flat direction $\phi_2$ can be realized as an axion which enjoys a perturbatively exact shift symmetry. Non-perturbative effects break this symmetry and generate a potential for $\phi_2$. However in the regime where the effective field theory is fully under control, stringy axions are expected to be ultra-light \cite{Svrcek:2006yi, Arvanitaki:2009fg, Cicoli:2012sz}, in agreement with our flat direction approximation.\footnote{In fact, the axion mass is set by the axion decay constant $f$ and the overall size of the axion potential $\Lambda$ as $m_{\phi_2}\sim \Lambda^2/f$. However in string compactifications these two quantities are not independent since $\Lambda$ is set by the instanton action $S$, $\Lambda^4 \sim M_{\mathrm{Pl}}^4\,e^{-S}$, which in many cases is inversely proportional to the axion decay constant $f$, i.e. $S\sim M_{\mathrm{Pl}}/f$ \cite{Svrcek:2006yi, Arvanitaki:2009fg, Cicoli:2012sz}. This implies that for $S\gg 1$ where the instanton expansion is under control, or equivalently for $f\ll M_{\mathrm{Pl}}$ as required by swampland bounds, the axion potential is in general very suppressed.}

Another interesting issue that we have not studied in detail is the coupling of $\phi_1$ and $\phi_2$ to matter. A direct coupling between one of the scalars and the barotropic fluid can have drastic effects on the late universe dynamics (see \cite{Bahamonde:2017ize} for a review of interacting dark energy models in the context of single-field quintessence). However if $\phi_1$ couples to ordinary matter with Planckian strength, fifth force bounds would require $\phi_1$ to be as heavy as at least $\mathcal{O}(1)$ meV ($\phi_2$, being a pseudo-scalar, would instead not suffer from this problem). It can be easily checked that this result can be achieved only by a huge tuning of the underlying parameter $k_2$ once the scale of scalar potential is required to be $V \sim 10^{-120}\,M_{\mathrm{Pl}}^4$. This implies that viable models would require a decoupling of $\phi_1$ from ordinary matter which could for example be realized via geometrical separation in the extra dimensions as in \cite{Cicoli:2012tz, Acharya:2018deu} (notice that $\phi_1$ could instead still have a direct coupling to dark matter).

Let us finally point out that quintessence models exacerbate the present tension between low-redshift data and the Planck determination of the Hubble constant based on $\Lambda$CDM \cite{Banerjee:2020xcn}. Given that at the level of the background our models are indistinguishable from standard quintessence, we expect them to feature the same $H_0$ tension. Let us however stress that the solution to this tension, if its observational evidence keeps gaining strength, does not need necessarily to come from the underlying dark energy model. Just to mention an example, we refer to extra dark radiation in the early universe \cite{Riess:2016jrr} which arises rather naturally in string models where relativistic ultra-light axions tend to be produced from the decay of heavy moduli \cite{Cicoli:2012aq, Cicoli:2015bpq}.

\end{document}